\documentclass{article}
\usepackage[utf8]{inputenc}
\usepackage[english]{babel}
\usepackage{authblk}
\usepackage{hyperref}

\title{ROOT for the HL-LHC: data format [2021-09-09]}

\author[1]{The ROOT Team\thanks{\href{mailto:rootdev@cern.ch}{rootdev@cern.ch}}: Axel Naumann}
\author[2]{Philippe Canal}
\author[1]{Enric Tejedor}
\author[1]{Enrico Guiraud}
\author[1]{Lorenzo Moneta}
\author[1]{Bertrand Bellenot}
\author[1]{Olivier Couet}
\author[3]{Alja Mrak Tadel}
\author[3]{Matevz Tadel}
\author[4]{Sergey Linev}
\author[1]{Javier Lopez Gomez}
\author[1]{Jonas Rembser}
\author[1]{Vincenzo Eduardo Padulano}
\author[1]{Jakob Blomer}
\author[1]{Jonas Hahnfeld}
\author[1]{Bernhard Manfred Gruber}
\author[5]{Vassil Vassilev}

\affil[1]{CERN, Geneva, Switzerland}
\affil[2]{Fermi National Accelerator Laboratory, Batavia, USA}
\affil[3]{University of California San Diego, USA}
\affil[4]{GSI, Darmstadt, Germany}
\affil[5]{Princeton University, USA}

\date{September 2021}

\begin{document}

\maketitle

This document discusses the state, roadmap, and risks of the foundational components of ROOT with respect to the experiments at the HL-LHC (Run 4 and beyond).
As foundational components, the document considers in particular the ROOT input/output (I/O) subsystem.
The current HEP I/O is based on the TFile container file format and the TTree binary event data format.
The work going into the new RNTuple event data format aims at superseding TTree, to make RNTuple the
production ROOT event data I/O that meets the requirements of Run 4 and beyond.

\section{Characteristics of the HEP I/O format}

ROOT provides the common file format for the data of all the LHC experiments, currently hosting more than 1\,EB of LHC data. 
WLCG estimates a yearly cost of 50\,MCHF for storage alone.
As such, a robust and efficient ROOT I/O is central to the success of the HL-LHC physics programme.
Improvements such as improved compression, faster reads and writes, and more reliable APIs benefit directly all the LHC experiments.

Given the projected volume of tens of exabytes and a life-time of HL-LHC data of several decades, it is natural and cost-effective to use a common I/O format that guarantees a lifetime compatible with HL-LHC.
A stable I/O format is also key to the long-term preservation of experimental results.
It is therefore crucial to retain expertise on I/O software and technology in the HEP community.

The different stages of the data flow---data taking or simulation, reconstruction, and analysis---come with different characteristics and requirements to the I/O.
Data taking puts a focus on fast writing and the ability to recover gracefully from unexpected premature termination of output streams. 
Reconstruction requires good interplay with multi-threaded frameworks, support for a rich event data model (EDM), and fast merging of multiple data streams from concurrent processing.
Analysis requires fast, selective reads both in terms of reading subsets of the events (``skimming'') and in terms of reading subsets of the collections and object properties in an event (``stripping'', ``slimming'' and ``thinning''). 
These requirements define a set of essential properties for the I/O and distinguish HEP-I/O from general
purpose I/O as provided by other libraries and formats.

\subsection{Essential Properties}

The ROOT I/O library aims at providing the following essential properties:

\begin{description}
\item[Robustness.] Protection against reading of corrupted data.
  This includes both protection against corruption on the storage or transfer level, and accidental misuse of the data access API.
\item[Stability.] In order to support long-term data preservation, the I/O library supports mechanisms for backward and forward compatibility. 
  This enables the evolution of its binary format (within limits) as well as schema evolution of the serialized objects: 
  ROOT I/O will translate for instance a jet read from an old file, to match the jet object in today's code.
  The I/O software development process prioritizes long-term maintenance and sustainable code design.
\item[Expressiveness.] HEP data are naturally represented as a large number of stochastically independent, structured events. 
  Events contain in turn collections of objects of (often significantly) varying sizes.
\item[Usability.] The I/O library is used both by software experts and by novice programmers, making reading and writing of data accessible to novice programmers. 
  In particular, arrays of C++ and Python objects in memory can be directly (de-)serialized to disk without an explicit schema definition by the user.
\item[Speed.] The columnar layout of the ROOT event data format and the ROOT I/O scheduler are tuned for the HEP use cases, especially for sparse reading and merging of data. 
  For these use cases, the ROOT I/O is faster than potential alternatives such as HDF5 or parquet~\cite{hepformats18}.
\item[Concurrency.] Data is commonly accessed in the context of concurrent applications. 
  The data format needs to be structured to facilitate inter- and intra-event parallelism. 
  The I/O layer itself can also deploy concurrent algorithms such as compression.
\item[Integration.] Data (de-)serialization is an integral part of the experiment computing workflows.
  The data format allows for storing experiment-specific and data management specific meta-data and
  integrates well with the experiment computing models.
\end{description}

The HL-LHC requirements on performance and efficient use of space in combination with the essential non-functional properties cannot be found in any single alternative format~\cite{hepformats18}.

\section{Overview of the I/O Components}

Several noteworthy components and classes collectively make the ROOT I/O library. 
Besides core components maintained by the ROOT team, the ROOT I/O library delegates functionality such as remote data access and authentication to the community's 3rd party implementations of choice.

\subsection{I/O Components in the ROOT Core Software}

The ROOT core software contains classes to read and write event data, as well as higher-level components for I/O scheduling, caching, and data reflection.

The TTree classes currently define the low level data format and data access API for LHC event data.
TTree uses a columnar on-disk layout.
It is used for all data product stages, from raw data to final-stage ntuples, and is the core property and benefit of ROOT's current I/O format, a data layout that is now adopted by many other ``big data'' tools.

The TFile container format allows to store TTree data, RNTuple data and non-event data such as histograms in a single file. The TFile format provides backward and forward compatibility over decades.

The RNTuple classes are the designated successor of the TTree classes.
Based on more than 25 years of experience with TTree and inspired by advances in I/O algorithms and hardware in industry and academia, RNTuple introduces a new data format and new data access APIs.
The format and the APIs are backwards incompatible to TTree.
This one-time compatibility break allows for the necessary flexibility in harnessing space savings, increasing the read/write speed, and improving the robustness of the API for the decades to come.
RNTuple includes an I/O scheduler that optimizes and parallelises the read requests for local devices (HDD, SSD) and remote storage (XRootD, HTTP, object stores, network file systems).
RNTuple includes the RNTupleLite library that provides low-level access to uncompressed data buffers through a C API and is independent of ROOT's core library.
The RNTuple binary format is specified to facilitate future 3rd party implementations, should the need arise.

ROOT's C++ and Python reflection capabilities are central to the I/O system; they enable serialization of user-defined objects. 
This type description is using ROOT's interpreter cling, which in turn utilizes the state-of-the-art C++ compiler libraries clang/llvm. 
Following clang's development for instance in support of new C++ standards allows cling and thus ROOT I/O to ``understand'' modern C++ code, not limiting the experiments in their expressiveness.
The serialization of events into the columnar on-disk layout effectively implements an AoS (array of struct) to SoA (struct of array) transformation at the I/O layer that is shared by all the RNTuple users.

Several higher-level I/O components provide support for specific use cases. 
In particular, ROOT provides support for application-defined data caching (TFilePrefetch) and dedicated support for multi-threaded and multi-node merging of partial results (TBufferMerger).

\subsection{Plugins and 3rd Party Components}

The ROOT core I/O integrates with the following 3rd party components that are important for the I/O of HL-LHC experiments.

\begin{itemize}
\item Remote access protocols: XRootD is critical for the production workflows; Davix is used for HTTP access and in particular for training, outreach and open science
\item Authentication plugins, namely X.509 certificates and SciTokens
\item Compression algorithms and lossy compression schemes; RNTuple and TTree provide the means to use both classes of compression
\item Connectors to efficiently accommodate new ``beyond file system'' storage (for instance Intel DAOS object store and Amazon S3 compliant cloud storage) and ``beyond grid'' compute resources such as exascale HPCs and clouds.
\end{itemize}

\section{Requirements from HL-LHC Experiments}

Compared to Run 1-3, we expect a tenfold increase in the number of events for HL-LHC.
While the exact effect of the increased data influx on file sizes and number of files is currently under investigation (see DOMA document), the file format needs to be prepared to handle significantly larger files and for a more flexible combination of physical files to virtual data sets.
That puts stronger robustness requirements on the I/O layer, e.g. for keeping data provenance information, for checksumming data and for the error handling of device failures.
In line with the experiment efforts on creating very compact analysis object formats, ROOT optimises the I/O performance especially for simple EDMs with structures of fundamental data types (``plain-old data'') and collections thereof~\cite{nanoaod19, physlite20}.

On the level of the computing model, remote I/O has already become a standard data access mode, a trend that will intensify towards HL-LHC (see also DOMA efforts).
HPC systems and clouds play an increasing role as resources beyond the standard grid.
These resources come with their own (remote and/or distributed) storage access technologies: cluster file systems and high-performance object stores for HPC sites, S3-like HTTP based object stores with URL based signatures and tokens for clouds. 
These storage types are expected to be increasingly used as temporary or tactical storage for LHC data.
In addition, some data is expected to be stored on a new, ``volatile'' storage class built from out-of-warranty drives for best use of existing resources, emphasising the need for rigid checksums on the file format level.

The increasing use of machine learning training and inference has an impact on storage, too.
The I/O system has to provide fast data pipes of event data to GPUs and other accelerators, ideally without involving the host CPU for the transfer.
Here, ROOT's columnar data is a perfect data layout for GPU acceleration.

\section{Evolution of the Technology Landscape}

The TTree I/O container was designed based on the hardware of the 1990s.
Meanwhile, the technology landscape has changed significantly, with generally greater architectural heterogeneity.
The boundaries between network, permanent storage, memory, and compute devices are blurring~\cite{storagehiera19}.
Remote storage is the norm rather than an exception, with varying degrees of remoteness and abstractions.
And compute devices are ever more parallel and specialised.

\subsection{Ultra-Fast Storage Devices}

Modern I/O devices such as 100\,GbE network cards and 10\,GB/s NVMe disks closely match the throughput of a CPU core.
As a result, I/O code paths must be highly CPU optimized.
In contrast to spinning disks with few platters, SSDs provide their full performance only when concurrently
accessing tens or hundreds of cells.
Furthermore, a new class of flash-based storage devices supports byte-addressed access as opposed to block-level access, which requires explicit support from the I/O layer to harness their full potential. Byte-addressable storage devices are already deployed for some performance-critical applications (e.\,g.~databases, HPC object stores) and may become part of the standard HEP storage hierarchy on the HL-LHC timescale.

\subsection{Parallelism and Accelerators}

TTree originally assumed single-core nodes, whereas HL-LHC's typical core counts will surpass 100 cores per box.
The I/O is now embedded in multi-threaded frameworks and makes active use of multithreading itself, for instance for data compression and decompression.
Standard environments increasingly deploy GPUs, used to offload for instance machine learning, tracking, and fitting tasks.
GPUs have a memory model different from CPUs and support new, reduced precision floating point layouts, which benefit from dedicated support by the I/O layer.

Some storage systems provide specific computational capabilities (``active storage''), e.g. for compression and checksumming~\cite{sds19}.
Active storage requires tight integration with the I/O application layer to use its capabilities.

\subsection{Distributed and Remote Storage}

Besides local and remote POSIX file system access, the I/O for HL-LHC needs to be prepared for file-less storage systems such as distributed object stores.
The scalability limits of the POSIX API make file-less storage systems ever more popular.
Cloud storage is already dominated by S3-like object stores, and HPC sites will likely rely on object stores for the next-generation supercomputers~\cite{hpcio17}.
With the divergence from the standard POSIX API and paradigms, however, a variety of data access APIs have emerged with no clear winner yet.
As a result, I/O software for HL-LHC needs to retain the flexibility to adapt, and a design that allows this adaptation with minimal cost, with respect to both performance and sustainability.

Even within the POSIX universe, applications can exploit an increasing number of tunable parameters by harnessing the actual file system's idiosyncrasies.
Ceph-FS for instance, which is used as a shared HPC cluster file system at CERN and elsewhere, can achieve improved performance by tuning custom file metadata.

\section{TTree Plans}

With the heritage of more than 1\,EB of LHC data, the TTree file format will remain supported in ROOT in parallel to the new RNTuple developments.
TTree also provides the baseline for RNTuple with respect to performance and functionality.
However, the priority in the TTree support will shift to long-term data preservation.
At the time scale of HL-LHC, TTree won't benefit from the increased and shared expertise and effort that RNTuple receives.

Tool support for disk-to-disk conversion from TTree to RNTuple is underway and expected for 2021. Large-scale conversion of data, e.\,g.~using spare I/O cycles, is conceivable according to experiments' needs as of the production release expected for 2024.
Alternatively, archived TTree datasets may be converted on the fly when reading from tape.

\section{RNTuple Potential and Roadmap}

The RNTuple I/O system is a multi-year effort to evolve to ROOT event data I/O for the challenges of the upcoming decades.
It comes with a rich R\&D programme on a compact data format, on performance engineering for modern storage hardware, on robust interfaces that are easy to use and hard to use incorrectly, and on various other I/O aspects~\cite{ntuple20, eprdet18}.
RNTuple is also an opportunity to grow the next generation of the community's long-term I/O experts.

RNTuple provides best performance for typical HEP use cases. 
It aims to closely match the I/O device speed, sustaining at least 10\,GB/s per box and 500\,MB/s per core for typical analysis throughput on the current generation of hardware and scaling up with the increased performance of new hardware.
To this end, RNTuple avoids virtual function calls and branches in hot code paths.
It is also designed for providing direct access to the I/O data buffers for zero-copy and bulk I/O.
On fast storage devices, RNTuple already shows 2--5 times better single-core performance than TTree~\cite{ntuple20}.
RNTuple is also designed for better scaling to high core counts.
The memory management is based on pages rather than on TTree ``baskets''.
Pages have a fixed size of $\mathcal{O}$(100\,kB) independent of the event sizes, which allows for more fine-grained parallelism and prevents a negative memory impact of large outlier events.

The RNTuple classes are conditionally thread safe (i.\,e., they use no globals or statics) and thus work well in multi-threaded applications.
Asynchronous, parallel I/O is the default.
RNTuple uses parallel range requests for remote storage (XRootD, HTTP) and local storage.
This approach optimally uses parallel storage devices such as SSDs and overlaps computation and data access.
Optionally, RNTuple connects to a task scheduler (e.g. Intel TBB) for transparent, task-parallel compression and decompression of pages.
The code is designed to allow for later offloading of compression to dedicated devices.

Compared to TTree, RNTuple has a significantly more compact data representation.
Comparisons show 15\,\% (lzma) -- 25\,\% (zstd) smaller files after compression with files from ATLAS, CMS, and ALICE.
The space savings are due to a very compact representation of (nested) collections and various smaller optimizations.
The introduction of the new RNTuple format also allows for changing the default compression algorithm from zlib to the better zstd.
Ongoing R\&D investigates lossy compression schemes; any benefits will come on top of the other savings.
Given the 50\,MCHF/year currently invested in WLCG storage, the community can benefit from considerable cost savings and higher luminosity exposed to analyses.

RNTuple has a 4-layered architecture: data access API, serialisation of C++ objects, management of storage-backed vectors of simple types, and the storage of byte ranges.
The layer separation facilitates adaptation to file-less storage systems~\cite{ntupledaos21}.
The byte range layer is prepared to work with byte-addressable as well as traditional block-based storage devices.
The explicit translation of C++ objects into arrays of simple types allows for robust data interpretability even in the absence of C++ reflection capabilities.
For instance, 3rd party tools can interpret the structure and (within limits) the meaning of RNTuple data without an understanding of C++ classes, as has been recently demonstrated in the context of research on a memory layout abstraction library~\cite{llamax21}.

The RNTuple API follows modern C++ core guidelines to provide a robust interface to users.
It uses smart pointers for a clear pointer ownership.
There are compile-time type-safe APIs for end-users and type-unsafe APIs with runtime checks for frameworks.
The systematic use of exceptions prevents silent I/O failures and silently corrupted data.
On the file format level, data and meta-data are systematically checksummed to guarantee data integrity against corruption due to storage devices or network transfers.

Like TTree, the programming model is designed for the needs of the HEP community, which is hard to achieve with industry standard tools. 
In particular:

\begin{itemize}
\item RNTuple is integrated as a data source with RDataFrame (see analysis chapter).
\item RNTuple provides a flexible meta-data API that allows experiments to store physics meta-data such as scale factors, as well as allowing data management tools like Rucio to access and write certain meta-data.
\item Like in TTree, vertical and horizontal data combinations are provided through friends and chains
\item Like TTree, RNTuple provides seamless support for serializing C++ and Python objects through ROOT's reflection facilities, as provided by ROOT's C++ interpreter cling
\end{itemize}

Having full control of the I/O layer allows for further integration in the future where needed.
Integration targets may include data access and authentication protocols, data management systems, Grid file transfer services, or tuning for specific file systems and network interconnects in HPC facilities.

\subsection{RNTuple Development Timeline}

The RNTuple development is a major undertaking of the ROOT team. We
continue to solicit community support and contributions. We foresee
several milestones along the way to validate the approaches and
coordinate integration with experiment frameworks. The following
timeline is conceived for a stable RNTuple developer work force of 2.5
experienced full-time engineers (see analysis chapter).

\begin{description}
\item[2018--2019:] Creation of the software architecture, R\&D on the binary file format, development of first prototypes
\item[2019--2020:] Integration in ROOT::Experimental, performance validations
\item[2021:]
  \begin{itemize}
  \item First integration in experiment frameworks (nanoAOD output module for CMSSW)
  \item Development of object storage backends (Intel DAOS, S3)
  \item Development of the schema evolution mechanisms
  \item Based on the experience of the prototypes, specification of the version 1 binary format
  \end{itemize}
\item[2022:]
  \begin{itemize}
  \item Optimization of RNTuple integration with ROOT's analysis interface RDataFrame ("bulk processing")
  \item Development of conversion tools, debug / inspection tools, facilitating the use and adoption of RNTuple
  \item Training of experiments' core developers
  \item Validation and of I/O type system completeness, in particular potentially missing STL types, opaque data types, references
  \end{itemize}
\item[2023:]
  \begin{itemize}
  \item Large-scale benchmarking and optimization of experiments' uses of RNTuple
  \item Development of automatic optimization features, such as branch-specific compression, prefetching and cache-reuse, page size, adjustment to storage-specific parameters (e.g. block size, queue depth)
  \item Validation of reading and writing data on PB scale
  \end{itemize}
\item[2024:] Training of physicists
\end{description}

\section{Adoption Work}

Besides the RNTuple development work itself, a successful community adoption requires additional development and training efforts.
This work cannot be done by the ROOT team alone. In particular, it requires investment from the experiments for the integration of RNTuple into their software frameworks.
To this end, integration efforts can be based on the experience with a first CMSSW integration for the generation of nanoAOD RNTuple files.

Furthermore, community support is required for training and dissemination, as well as the validation and optimization of the RNTuple I/O to cover the wide spectrum of usage contexts of ROOT I/O.

\section{Risks}

For the efforts to prepare the ROOT foundations for HL-LHC, we see the following challenges:

\begin{enumerate}
\item Keeping the schedule of the RNTuple implementation plan which depends on the availability of the required development resources.\\
  The potential impact is a fragmentation of the community using several, ad-hoc I/O approaches (e.g. TTree, protobuf, etc.) and along with it the risk of increased storage needs and reduced compute efficiency.
  As a risk mitigation, we foresee the gradual rollout of RNTuple, starting from derived data products up to raw data. 
  This allows directing development efforts to critical areas in an agile way, based on feedback.
\item Long-term retention of TTree and RNTuple I/O experts.\\
  The potential impact is a trust erosion in the data format and inefficiencies due to workarounds. 
  As a mitigation we consider a thorough code development and documentation process according to best practices as well as education and sharing of expertise through R\&D and adoption. 
  We also consider the existing permanent positions invested in I/O an important risk mitigation.
\item Optimal design of the RNTuple format and API for the hardware and software requirements for Run 4.\\
    The potential impact is a loss of performance, thus a limitation on the efficiency of HL-LHC computing workflows, and in the worst case partial data loss.
    Other risk dimensions include the lack of flexibility needed to address future challenges or evolving requirements, or unacceptable limitations of supported experiments' data formats. 
    As a mitigation, we plan for involvement of the stakeholders in the RNTuple specification, validation tests with experiments, and best practices used during the development. 
    The RNTuple design is informed by the TTree experience, allowing for adaptation in areas expected to change on the time scale of HL-LHC, such as object store technology and active storage.
\item Availability of remote access protocol libraries.\\
  The availability of a well-maintained XRootD client library and its integration in ROOT is critical for the HEP community. 
  The availability of a well-maintained Davix client library and its integration in ROOT is important for training, outreach, and open science.
  While not directly related to ROOT, HTTP library support is also critical for the HEP community for 3rd-party copies (storage-to-storage transfers).
\item Optimization of experiments' persistent data model for RNTuple.\\
  The RNTuple design anticipates the ongoing optimization of the experiments' persistent data model. 
  Investing in developer FTEs on both ROOT's and the experiments' side to ensure a close feedback loop between the ROOT team and experiments and for the early adoption by experiments is critical.
  We believe that benefits provided by RNTuple are sufficiently convincing to warrant the experiments' transition
  with high priority.
\item Evolving ROOT reflection support.\\
  To accommodate the experiments' continuous demand for the always newest C++ standard, with its additional features, efficiency, and expressiveness, ROOT needs to have the ability to understand newest C++ code, to extract type description (reflection), and to serialize objects of these types.
  This requires excellent C++ support as provided by one of the few industry-grade C++ compilers and libraries, clang.
  Cling and ROOT must follow and ideally influence clang's steady yet fast-paced evolution, which requires very specialized expertise and dedicated, continuous effort.
  Expert effort is also required for evolving and supporting ROOT's type description system, and for providing stable interfaces in ROOT despite an ever changing clang, for the lifetime of HL-LHC's usage of C++.
\end{enumerate}

\bibliographystyle{unsrt}
\bibliography{bibliography}

\end{document}